\documentclass[a4paper,11pt]{article}
\pdfoutput=1 

\usepackage{jcappub} 

\usepackage[T1]{fontenc} 

\usepackage{graphicx}
\usepackage{subcaption}
\usepackage{amsmath} 
\usepackage{verbatim} 
\usepackage{mathrsfs}
\usepackage{color}
\usepackage{amsfonts}
\usepackage{graphics}
\usepackage{epsfig}
\usepackage{hyperref}
\usepackage{caption}
\newcommand{\x}{\textbf{x}}

\newcommand{\mH}{\mathcal{H}}

\newcommand{\mM}{\mathcal{M}}
\newcommand{\mN}{\mathcal{N}}
\newcommand{\mL}{\mathcal{L}}
\newcommand{\nn}{\nonumber \\}

\usepackage{braket}
\newcommand{\be}{\begin{equation}}
	\newcommand{\ba}{\begin{eqnarray}}
		\newcommand{\n}{\nonumber}
		\newcommand{\ee}{\end{equation}}
	\newcommand{\ea}{\end{eqnarray}}
\DeclareFontFamily{U}{rsfs}{}         
\DeclareFontShape{U}{rsfs}{m}{n}{<5> rsfs5 <6><7> rsfs7          %
	<8><9><10><10.95><12><14.4><17.28><20.74><24.88> rsfs10}{}     %
\DeclareMathAlphabet{\mathfs}{U}{rsfs}{m}{n}                     %
\newcommand{\mfs}[1]{\mathfs {#1}}

\newcommand{\sM}{{\mfs M}}

\title{A clarification on prevailing misconceptions in unimodular gravity}

\author[a]{Gabriel R. Bengochea,}

\author[b,c]{Gabriel Le\'{o}n,}

\author[d]{Alejandro Perez,}

\author[e]{and Daniel Sudarsky}

\affiliation[a]{Instituto de Astronom\'\i a y F\'\i sica del Espacio (IAFE), CONICET - Universidad de Buenos Aires, (1428) Buenos Aires, Argentina}
\affiliation[b]{Grupo de Cosmolog\'{\i}a, Facultad de Ciencias Astron\'{o}micas y Geof\'{\i}sicas, Universidad Nacional de La Plata, Paseo del Bosque S/N 1900 La Plata, Argentina}
\affiliation[c]{CONICET, Godoy Cruz 2290, 1425 Ciudad Aut\'onoma de Buenos Aires, Argentina}
\affiliation[d]{Aix Marseille Univ, Universit\'e de Toulon, CNRS, CPT, Marseille, France}
\affiliation[e]{Departamento de Gravitaci\'{o}n y Teor\'{\i}a de Campos, Instituto de Ciencias Nucleares, Universidad Nacional Aut\'{o}noma de M\'{e}xico, A. Postal 70-543, M\'{e}xico City 04510, M\'{e}xico}

\emailAdd{gabriel@iafe.uba.ar}
\emailAdd{gleon@fcaglp.unlp.edu.ar}
\emailAdd{perez@cpt.univ-mrs.fr}
\emailAdd{sudarsky@nucleares.unam.mx}

\abstract{ The traditional presentation of Unimodular Gravity (UG) consists on indicating that it is an alternative theory of gravity that restricts the generic diffeomorphism invariance of General Relativity. In particular, as often encountered in the literature, unlike General Relativity, Unimodular Gravity is invariant solely under volume-preserving diffeomorphisms. That characterization of UG has led to some confusion and incorrect statements in various treatments on the subject. For instance, sometimes it is claimed (mistakenly) that only spacetime metrics such that $|$det $g_{\mu \nu}| = 1$ can be considered as valid solutions of the theory. Additionally, that same (incorrect) statement is often invoked to  argue that some particular gauges (e.g. the Newtonian or synchronous gauge) are not allowed when dealing with cosmological perturbation theory in UG. The present article is devoted to clarify those and other misconceptions regarding the notion of {\it diffeomorphism invariance}, in general, and its usage in the context of UG,  in particular. }

\begin{document}
\maketitle
\flushbottom

\section{Introduction}

Diffeomorphism invariance is a notion that, despite its ubiquitous occurrence in contemporary physics, and its apparent simplicity, is often the source of some serious misunderstandings. The consideration of theories that are said to be less than fully diffeomorphism invariant has, not surprisingly, generated even more propensity to confusion. One such example, that has recently attracted some attention, is the case of the so called \emph{Unimodular Gravity} (UG) (e.g. \cite{UG0,UG1,UG2,UG3,UG4,ellis2010,Gao2014,ellis2015,Basak2015,Nojiri2015,Nojiri2016,Nassur2016,Bamba2016,Daouda2018,Bonder18,Garcia2019,Astorga2019,Corral19,GUG2,Corral20,Nucamendi20,deCesare2021,Fabris2021,Fabris2021GW,GUG1,Landau2022,Bonder22,Pia2023,Alvarenga2023,Maroto2023,Nucamendi2023}, see also \cite{Carballo2022} for a recent review). The purpose of this manuscript is to attempt to clarify some of the confusions and misconceptions that appear in several of the works mentioned.

Let us start by noting that a vast number of physical theories are currently formulated using the language of differential geometry, and, thus, rely on the notion of differential manifold $\sM$ as a starting point. Moreover, these theories are often specified in terms of an action functional, that is given by an integral over such a manifold involving the dynamical fields occurring in the theory, taken to be represented by suitable tensor or spinor fields  over the manifold. The first point  to note is that the only notion of integration over an $n$-dimensional manifold that is mathematically well defined (given the very notion of what a differentiable manifold is), is the integral of an $n$-form. Such integrals are, by construction, always diffeomorphism invariant. Thus, the theories we have been talking about are always, in that sense, diffeomorphism invariant.

If an action principle depends on fields $\left\{\psi_A \right\}_{A=1}^N$,  then it is convenient to separate the fields into two classes:
dynamical fields, which we choose as the first set $\left\{\psi_A \right\}_{A=1}^{N_d\le N}$, and background fields $\left\{\psi_A \right\}_{A=N_d+1}^{N}$, and adapt the notation so that such distinction is apparent. We hence denote the action
\be\label{unos}
S_{\left\{\psi_A \right\}_{A=N_d+1}^N}\left[\left\{\psi_A \right\}_{A=1}^{N_d}\right]
\ee
by placing background fields as subindices. Dynamical fields are placed as arguments (encased in square brackets), and are those that are varied in the definition of the field equations, in the action, namely
\be\label{dos}
\frac{\delta S}{\delta \psi_A}=0, \ \ \ 1 \le A \le N_d.
\ee
Background fields are not varied and are there as complementary structures necessary to express the action principle as an integral of an $n$-form in an $n$-dimensional manifold.

Any well defined field theoretical action principle satisfies the following condition that we call {\em tautological diffeomorphism invariance}, namely
\ba \label{uno}
&& S_{\left\{\psi_A \right\}_{A=N_d+1}^N}\left[\left\{\psi_A \right\}_{A=1}^{N_d}\right] =\\ \n
&& \ \ \ \ \ \ \ \ \ \ = S_{\left\{\phi(\psi_A) \right\}_{A=N_d+1}^N}\left[\left\{\phi(\psi_A) \right\}_{A=1}^{N_d}\right],
\ea
where $\phi(\psi)$ denotes a diffeomorphism $\phi:\sM\to \sM$.
The previous invariance is simply associated to the fact that coordinates are mere labels of points in the manifold, and, as such, completely
arbitrary in specifying the action principle.
A key property of {\em coordinate covariance}, equation \eqref{uno} is that  the field equations \eqref{dos}, can be formulated in arbitrary coordinates. This fact is trivial and should not be confused with {\em dynamical diffeomorphism invariance}. A theory will be said to be {\em  dynamical diffeomorphism invariant} if there are no background fields, i.e., if $N_d=N$. The emblematic example of  a dynamical diffeomorphism invariant theory is General Relativity where metric and matter fields are all dynamical (no background fields are present); for instance pure gravity in four dimensions
\be\label{gr}
S[g_{ab}]=\frac{1}{16 \pi G} \int_{\sM} \sqrt{|g|} R[g_{ab}] dx^4.
\ee
When coupling General Relativity to matter fields that involve fermions and gauge fields, it is necessary to introduce further mathematical structures (fiber bundles, spin structures, etc) in order to maintain the necessary tautological diffeomorphism invariance that makes the action well defined. For example, this is the case with Cartan's formulation of General Relativity, which provides the customary formulation for the coupling of gravity with the standard model of particle physics. Other celebrated examples include Chern-Simons theory in three dimensions, BF theory in any dimensions, as well as General Relativity in arbitrary dimensions.

As mentioned, condition \eqref{uno} universally implies that the field equations are covariant: they can be expressed in arbitrary coordinate systems once all fields are suitably transformed. However, condition \eqref{uno}, has a dramatic dynamical implication for dynamical diffeomorphism invariant theories: in the absence of background fields, the field equations can determine the dynamics of fields only up to diffeomorphism. Two solutions differing by a diffeomorphism are to be taken as physically equivalent, thus the transformation $\psi_A\to \phi(\psi_A)$ has to be interpreted as a gauge transformation. This is best seen in the context of the initial value formulation of such field theories, most clearly in their Hamiltonian description using Dirac's theory of gauge symmetries.

An example of a theory satisfying \eqref{uno} but not being dynamical diffeomorphism invariant is electromagnetism in 4 dimensions, on a given background metric, whose action is
\be\label{em}
S_g[A]=-\frac 14\int_{\sM} \sqrt{|g|} g^{ab}g^{cd} (dA)_{ac}(dA)_{bd} \ dx^4.
\ee
Electromagnetism is not dynamical diffeomorphism invariant due to the presence of a non trivial non-dynamical field: the metric $g_{ab}$.
The field equations $\delta_A S=0$ are the standard Maxwell equations on an arbitrary metric background, which, as a consequence of \eqref{uno}, can be expressed in arbitrary coordinates. A particularly simple example is the case where $g_{ab}=\eta_{ab}$ is the flat Minkowski metric. In that case, it is some times convenient to express Maxwell equations in terms of inertial coordinates. However, we can equally well express the field equations in whatever coordinate system that might fit the physics we intend to describe. Condition \eqref{uno} implies that, as long as all fields (background and dynamical) are  transformed to the new coordinates, the equations remain covariant. The fact that electromagnetism on a fixed background breaks dynamical diffeomorphism invariance does not restrict the physicist from using different coordinates to analyse the physics.

The distinction between dynamical or non dynamical fields is a matter of the particular application. For instance, in the previous case, the metric is taken as a non-dynamical background field when describing electromagnetic phenomena on a given spacetime geometry while neglecting their possible gravitational effects. However, the very same differential geometry expression \eqref{em} would be used when coupling electromagnetism to gravity while now promoting the metric $g_{ab}$ from background to dynamical field, with the suitable addition of \eqref{gr} in the expression of the  action of Einstein-Maxwell theory, namely
\ba \label{emgr}&&
S[g_{ab},A_a]=\frac{1}{16 \pi G}\int_{\sM} \sqrt{|g|} R[g_{ab}] dx^4 \\ \n && \ \ \ \ \ \ \ \ + \int_{\sM} \sqrt{|g|} g^{ab}g^{cd} (dA)_{ac}(dA)_{bd} \ dx^4.
\ea

A similar and related fact occurs with the notion of special relativistic covariance, the underlying principle of Special Relativity (SR), providing for the equivalence of all inertial frames (and the coordinates that one might associate to them). Take, for instance, a proposal of violation of SR that was popular a couple of decades ago,  that took as basic  hypothesis the idea that, as a result of a quantum gravity granularity of spacetime, the dispersion relation of free particles would be modified as $E^2= {\vec P}^2 + m^2 +  \xi E^3/ M_{P}$ (with the Planck mass $M_{P}$ indicating the quantum gravitational nature of the effect, and $\xi$ an unknown parameter on which researchers diligently work to set  bounds). The equation clearly was not covariant in the special relativistic sense, but could nevertheless be written in any terms of coordinates adapted to any inertial frame. Once one recognized that feature, such a proposal implied the presence of an additional geometric structure in the otherwise Minkowski spacetime. That structure is, in this case,  a global  vector field $W^a$ (which for simplicity can be taken as a constant field having vanishing covariant derivatives), representing a preferential frame in which the dispersion relation took the given form. The point is that the dispersion relation could now be written in a special-relativistic covariant-looking form as: $P^aP_a + m^2 + (\xi/ M_{P})(P^a W_a)^3=0$. The price of violating the special relativistic version of covariance is not that one might not use other coordinates, but simply that the novel geometrical structure (in this case the vector field $W^a$) would occur explicitly in the equations.

One of the points we want to emphasize in this manuscript is the direct generalization of this lesson; both in general terms, and, in particular, for the case of UG. The presence of  some additional geometric structure in the theory (which in the case of UG is a fixed ``non-dynamical'' 4-volume element, besides the one associated to the space-time metric) in no way limits the possibility of working with the theory in any coordinate chart: UG satisfies equation \eqref{uno}.  Moreover, as we will see, the fact that, in order to specify a background four volume structure, one often needs to make use of some coordinate chart, has the consequence that, in practice, one might work in any coordinate chart without even paying the price of having the additional structure occurring in the equations one needs to deal with. In the case of UG, this means that there is no restriction in the coordinates one might use to work with the theory. Moreover, even when this is not explicitly obvious, UG is almost generally covariant, in the sense that it preserves all the dynamical gauge symmetries of General Relativity, with the exception of a single (among the infinitely many local gauge symmetry generators) global generator that is broken \cite{Henneaux:1989zc, UG2}. This means that the dynamical gauge structure  and hence the possibilities of gauge choices one can make use of in UG, for instance in working in perturbative treatments is just the same as the one in GR.

Another point that is worth discussing concerns precisely the characterization of an object as dynamical or non-dynamical. Here, the issue is that such a distinction need not be one of principle, and can, in fact, vary from situation to situation. One might, for instance, take an external electromagnetic field acting on a set of charged particles as  a  non-dynamical entity, for the purpose of the analysis of certain situation (to a desired degree of approximation), while acknowledging that such an entity is, in principle, a fully dynamical object whose equations of motion are not being taken into account in the given context. So, it is the use or lack of use of the corresponding dynamical equations what turns an object, within a certain treatment, into  a dynamical or non-dynamical one [e.g. recall the previous discussion comparing the action \eqref{em} and \eqref{emgr}].

In fact, one might even consider situations in which one acknowledges that such equations are not at the time known, or that one will work with approximations that ignore certain aspects. Furthermore, when considering a problem, one might treat certain objects in a classical manner, implying that one will consider such an object as ruled by the classical equations of motion, or,  on the contrary, imagine some effective equations that take into account certain quantum aspects, or unknown features of the problem.
As a simple example, we can imagine deriving the equation of motion of a free scalar field, and then adding to it some  effective friction terms, which reflect the fact that the field is interacting with certain unknown degrees of freedom and is  dissipating energy into the corresponding channel. Moreover, one might consider representing unknown effects, such as aspects one is considering to be connected with quantum features of the problem, or even with quantum gravity, and thus include effective descriptions involving deviation from the classical equations of motion.

The next issue we want to clarify concerns the implications of issues such as energy conservation, or the possible violation of that, in the context of both GR and UG. As we will see, the standard notion of diffeomorphism invariance, as employed in General Relativity, implies that the classical equations of motion for the geometrical variables are consistent only if the matter fields are, in turn, characterized by equations that strictly enforce the conservation of the energy-momentum. That is, when considering the classical equations for the metric, the characterization of the matter fields, even if not relying on the strict classical equations for them, must be such that energy-momentum is conserved $\nabla^aT_{ab}=0$ (for instance, one might use  semiclassical equations as long as the expectation value of the energy-momentum tensor is conserved).
The gauge structure of UG allows for violations of energy-momentum conservation $\nabla^aT_{ab}\not=0$, with the restriction that the new continuity equation must hold: $\nabla^aT_{ab}-\nabla_b \Lambda=0$ (energy lost or gained in the matter sector leaks into, or comes from, the dark energy component).

The paper is organized as follows: we begin in section \ref{Sec_UGeqs} reviewing Unimodular Gravity from a variational principle and mentioning confusions regarding the diffeomorphism invariance of the theory. In section \ref{Sec_UGsol}, we address the issue of the choice of coordinates, and how it tends to generate misunderstandings when working with UG. Then, in section \ref{Sec_perturb}, we make a concise review of basic aspects of covariant perturbation theory. In section \ref{Sec_UGpert}, we focus on the cosmological perturbation theory in UG. There, we will discuss how, in agreement with \cite{Henneaux:1989zc, UG2} and contrary to a widespread set of confused statements, there are no differences with respect to standard General Relativity, when implementing the perturbation theory in Unimodular Gravity . Finally, in section \ref{conclusions}, we present our conclusions.

Throughout this work, we will use a $(-,+,+,+)$ signature for the spacetime metric and units where $c=1$. Also, we will follow Wald's convention and notation for the geometrical objects \cite{Waldbook}. Particularly, we will use  the \textit{abstract index notation} which makes a distinction between index notation and component notation. For example, $T_{ab}$ denotes a tensor $T$ of type (0,2), namely it acts on two vectors. The latin indices act as reminders of the number and type of variables the tensor acts on.  On the other hand, greek indices denote components of a tensor in a given basis. Thus, for instance $T_{\mu \nu}$  will denote the  component of the tensor $T_{ab}$ in  some   particular coordinates. In addition, we will follow Wald's convention for the Riemann tensor  $R_{abc}^{\: \: \: \: \: \: d} w_d \equiv (\nabla_a \nabla_b -  \nabla_b \nabla_a ) w_c$.

\section{Unimodular gravity from a variational principle}\label{Sec_UGeqs}
In this section, we revise, for a pedagogical reason, the classical equations of UG using the variational principle.
After that, we will point out the potential sources of confusion regarding the diffeomorphism invariance of the theory, and its relation with a possible non-conservation of the energy-momentum tensor.
The UG action can be expressed through the functional
\begin{eqnarray}\label{accion0}
	S_{\varepsilon}[g^{ab}, \Psi_M;\lambda] &=& \frac{1}{2 \kappa} \int \left[  R \epsilon_{abcd}^{(g)}  -2 \lambda (\epsilon_{abcd}^{(g)}  - \varepsilon_{abcd} ) \right]  \nn
	&+& \int  \mL_{M}[g^{ab}, \Psi_M] \epsilon_{abcd}^{(g)},
\end{eqnarray}
where $\kappa \equiv 8 \pi G$, $R$ is the Ricci scalar, $\varepsilon_{abcd}$ is a fiduciary 4-volume element (supposed to be supplied by the theory), and $\epsilon_{abcd}^{(g)} $ is the 4-volume element associated to the metric $g_{ab}$. The scalar $\lambda(x)$ is a Lagrange multiplier function, and $\mL_M$ is the Lagrangian density of the matter fields represented by $\Psi_M$. We start by recalling that any two 4-forms in 4 dimensions are related by a real valued function $h$, which, in the case  of the volume forms involved in our discussion, implies that
\be \epsilon_{abcd}^{(g)}  =  h \varepsilon_{abcd}.\ee
In fact, as we already noted, if we choose to use some specific coordinates we can write \be \epsilon_{abcd}^{(g)} =\sqrt{-g}  dx^0_a \wedge dx^1_b  \wedge dx^2_c  \wedge dx^3_d,\ee   and \be \varepsilon_{abcd} = f dx^0_a \wedge dx^1_b  \wedge dx^2_c  \wedge dx^3_d,\ee
so then $\sqrt{-g} = hf $\footnote{Notice that one often finds the \emph{``unimodular constraint''} erroneously expressed as the demand that $h=1$, or as $\sqrt{-g} = 1$. It is true that one might chose to work with coordinates where that holds, however, as we will see, it is a mistake to think that the theory requires that. We will say more about this in the next section.}.
The classical equations of motion are obtained by requiring the extremization of the total action, when considering variations of Eq. \eqref{accion0} with respect to the dynamical variables: $g^{ab}$, $\lambda$ and $\Psi_M$. This procedure yields:
\begin{equation}\label{EFE}
	R_{ab} - \frac{R}{2}g_{ab} + \lambda(x) g_{ab} = \kappa T_{ab}
\end{equation}
\begin{equation}\label{constraint0}
	\epsilon_{abcd}^{(g)}  = \varepsilon_{abcd}
\end{equation}
\begin{equation}\label{KG}
	\frac{\delta S^M}{\delta \Psi_M} = 0.
\end{equation}
Additionally, we have used the standard definition:
\begin{equation}\label{defTab}
	\delta_g S^M [g^{ab}, \Psi_M] = -\frac12\int   \epsilon_{abcd}^{(g)}T_{lm} {\delta g^{lm}}
\end{equation}
of the energy-momentum tensor, in terms of the matter action functional derivative with respect to the metric. Equation \eqref{KG} is the field equation for the matter fields.
We can eliminate the Lagrange multiplier from Eq. \eqref{EFE}. Taking the trace of such an equation results in
\begin{equation}\label{lambda}
	\lambda = \frac{\kappa T + R}{4},
\end{equation}
where $T = g^{ab} T_{ab}$ is the trace of the energy-momentum tensor.  Substituting the former expression in Eq. \eqref{EFE}, leads to the trace-free part of Einstein's field equations, namely
\begin{equation}\label{UGecs}
	R_{ab} - \frac{1}{4}  g_{ab} R  =   \kappa \left( T_{ab} - \frac{1}{4}  g_{ab} T \right)
\end{equation}
which are the UG equations for the gravitational field.

\subsection{Energy-momentum conservation in UG}\label{Sec21}

Let us explore a consequence of the invariance of the matter action under the limited set of infinitesimal \emph{``volume preserving diffeomorphisms''} $\xi$  (despite the fact that we take it to be fully dynamical diffeomorphism invariant). Setting to zero the variation of the matter action $S^M$, one obtains
\begin{equation}\label{varSM}
	0 = \delta_{\xi}S^M = \int \epsilon_{abcd}^{(g)}
	\:  \frac{\delta S^M}{\delta g^{lm}} \delta_{\xi} g^{lm} + \int \epsilon_{abcd}^{(g)}
	\:  \frac{\delta S^M}{\delta \Psi_M} {\delta}_{\xi} {\Psi_M},
\end{equation}
Assuming that $\Psi_M$ satisfies the matter field equations, then $\delta S^M/ \delta \Psi_M |_{\Psi_M} = 0$  and the last term in Eq. \eqref{varSM} makes no contribution. Note that one might at this point, consider  instead, that something that is not being described at the level of the action is happening regarding the evolution matter fields, and that, therefore, these fields do not satisfy their classical equations of motion. In that case, we can either ignore their variation, or take the spacetime characterization of the matter fields as fixed, and not varying under the diffeomorphism (like the example of the external electric field discussed above). We will touch back on this point below.

When we take the matter fields in spacetime as given, while we restrict our consideration to volume preserving diffeomorphisms, the requirement that the matter action be invariant will lead to some constraints. The point is that, in such  case, one is led to focus on the corresponding infinitesimal generators represented by vector fields $\xi^a$ with vanishing divergence, i.e. $\nabla_a \xi^a = 0$. The general (local) form of those equations is given by\footnote{The total antisymmetric tensor $\epsilon^{abcd}$ is defined through the condition $\epsilon^{abcd} \epsilon_{abcd} = - 4!$, where $\epsilon_{abcd}$ is the Levi-Civita tensor, which, in fact,  is proportional to $\epsilon_{abcd}^{(g)} $.}  $\xi^a= \epsilon^{abcd} \nabla_b\omega_{cd}$, where the two-form $\omega_{ab}$ is arbitrary. Therefore, the variation $\delta g^{ab}$ (corresponding to volume preserving diffeomorphisms) has the form $\pounds_\xi  g^{ab} = -2 \nabla^{(a} \xi^{b)}$ with $\xi^a= \epsilon^{abcd} \nabla_b\omega_{cd}$. Taking into account all these previous elements, and the definition of $T_{ab}$ \eqref{defTab}, the variation \eqref{varSM} is
\begin{eqnarray}\label{varSM2}
	0 &=& \delta_\xi S^M = \int \epsilon_{abcd}^{(g)}   \: T_{lm} \nabla^{l} \xi^m \nn
	& = &- \int  \epsilon_{abcd}^{(g)} \: (\nabla^l T_{lm}) \xi^m \nn
	& = &  \int  \epsilon_{abcd}^{(g)}    \:  \nabla_n  (\nabla^l T_{lm}  \epsilon^{mnks} ) \omega_{ks},
\end{eqnarray}
where we have integrated by parts two times and assumed that the total derivative terms vanish at infinity. Introducing the definition $J_a \equiv \nabla^b T_{ab}$, the condition \eqref{varSM2} implies that $\mathrm{d} J =0$ (since $\omega_{de}$ is locally arbitrary), and, therefore, we must have
\begin{equation}\label{defQ}
	J_a = \nabla_a Q,
\end{equation}
for some scalar field $Q$. Thus, we have the equation
\begin{equation}\label{conservTab}
	\nabla^a (T_{ab} - g_{ab} Q) =0.
\end{equation}

Additionally, applying $\nabla^a$ to both sides of Eq. \eqref{EFE}, using Eq. \eqref{conservTab}, and making use of Bianchi's identities $\nabla^a G_{ab} =0$, one finds
\begin{equation}\label{lambdax}
	\lambda(x) = \Lambda_0 + \kappa Q(x),
\end{equation}
where $\Lambda_0$ is simply a constant of integration. In fact, if $Q =$ constant, the standard conservation law for $T_{ab}$ is restored.

In other words, in UG the energy-momentum conservation might be taken as an additional assumption (and it is often done) imposed on the theory (when one, for instance, considers an energy-momentum that is given externally, or that of a fluid for which one does not have the appropriate dynamical evolution laws). Nevertheless, in principle, one might choose not to adopt such an assumption, in which case, the important fact is that UG generically admits a violation of the energy-momentum conservation, as long as the 1-form $J$ is closed. It is for this last case that UG introduces deviations from GR.

Here, we must note that, for the usual case of dynamical diffeomorphism invariant theories, the classical equations of motion of the matter fields in question do imply, by themselves, the conservation of the energy-momentum tensor; but we should emphasize that such a requirement applies just to the classical equations of motion, and one might be interested in considering the inclusion of novel effects (which could be of purely quantum mechanical origin, such as spontaneous collapse of the quantum state \cite{Bassi2,sudarskyPRL1}, and referred to a semiclassical treatment where the energy-momentum tensor is replaced by its expectation value), or effects of unknown degrees of freedom (possibly arising from quantum gravity) related to, say, some granularity of spacetime, and which are not reflected by the classical equations of motion.

\subsection{An alternative derivation}\label{Sec22}

The generic restriction on the non-conservation of the energy-momentum tensor in UG, characterized by Eq. \eqref{conservTab}, was derived by restricting consideration to the volume preserving diffeomorphisms when performing the variation of $S^M$.  And while such a derivation is valid within UG, there is a slightly different derivation that does not involve such limitations. The important feature of  this alternative approach is that one starts by considering the relevant variation of the action \eqref{accion0} as involving \emph{all the geometrical objects, i.e. by considering diffeomorphisms acting both}, on the dynamical variables, and also on the non-dynamical ones (in the sense of what we termed tautological diffeomorphism invariance expressed in Eq. \eqref{uno}). The action \eqref{accion0} is the sum of the following three terms:
\begin{equation}\label{acciongrav}
	S^{\rm EH}[g^{ab}] \equiv	\frac{1}{2 \kappa} \int  R \epsilon_{abcd}^{(g)},
\end{equation}
\begin{equation}\label{accionmatt}
	S^M[g^{ab}, \Psi_M] \equiv	\int  \mL_{M}[g^{ab}, \Psi_M] \epsilon_{abcd}^{(g)},
\end{equation}
and
\begin{equation}\label{accionf}
	S_\varepsilon[g^{ab}; \lambda] \equiv	\frac{-1}{\kappa} \int \lambda (\epsilon_{abcd}^{(g)}  - \varepsilon_{abcd} ).
\end{equation}

Due to \eqref{uno}, the variation under a general diffeomorphism of each and every term in the previous list vanishes independently. We concentrate on the Einstein-Hilbert term, and we find
\begin{eqnarray}\label{varS2}
	0 &=& \delta_\xi S^{\rm EH} \\
	\nonumber &=&  \frac{1}{\kappa} \int  \epsilon_{abcd}^{(g)}   \left(-R_{lm}  + \frac{R}{2} g_{lm}  \right)  \nabla^l \xi^m
\end{eqnarray}
Making use  of the UG field equations \eqref{EFE}, we obtain
\begin{equation}\label{varS2b}
	0 = \delta_\xi S^{\rm EH} =	 \int \epsilon_{abcd}^{(g)} \left(  \kappa^{-1} \lambda g_{lm} - T_{lm}   \right) \nabla^l \xi^m,
\end{equation}
then, defining $J_a \equiv  \nabla^b T_{ab}$, and integrating by parts (neglecting boundary terms that vanish due to suitable boundary conditions) yield
\begin{equation}\label{varS2c}
	0 = \delta_\xi S^{\rm EH} 	= - \int \epsilon_{abcd}^{(g)} \xi^m \left(    \nabla_m \kappa^{-1} \lambda - J_m  \right).
\end{equation}
The above equation is now valid for arbitrary $\xi$ and thus, from Eq. \eqref{varS2c}, we find that the 1-form $J$ is such that $J= \kappa^{-1}  \mathrm{d} \lambda$, therefore, $\mathrm{d}J =0$, or locally $J_a = \nabla_aQ$. Namely, we recover Eq. \eqref{conservTab}, which reflects the generic non-conservation of $T_{ab}$.

\section{On the solutions of unimodular gravity equations}\label{Sec_UGsol}

Here, we consider the subject of actually solving the UG equations for a given physical situation. This point raises delicate and subtle issues that we will address in the present section. The idea is that, in general, to find a solution to the field equations it is necessary to use specific coordinates, and the issue of their choice tends to generate misunderstandings when working with UG.

\subsection{Solutions just by ``changing coordinates"?}\label{Sec_UGsolA}

We begin by recalling that there are two equations to be solved from the gravitational side of the theory. These are Eqs. \eqref{constraint0} and \eqref{UGecs}, which, for convenience,  we write them again in the following:
\begin{equation}\label{UGecs2}
	R_{ab} - \frac{1}{4}  g_{ab} R  =   \kappa \left( T_{ab} - \frac{1}{4}  g_{ab} T \right)
\end{equation}
and
\begin{equation}\label{constraint}
	\epsilon_{abcd}^{(g)}  = \varepsilon_{abcd},
\end{equation}
also, we remind the reader that $\varepsilon_{abcd}$ is taken as given. Here, we note that, in principle, for each given $\varepsilon_{abcd}$, one would have a different theory (or a different setting of external conditions), in the sense that one would obtain different solutions in each case.

Next, let us focus on the issue of finding solutions of the theory, particularly, on the role of the constraint on the 4-volume. To do so, let us assume that we are working with a fixed set of coordinates on the spacetime manifold $\mM$, and call them $y^\mu$. Also, assume we are given a fiduciary 4-volume as
\begin{equation}\label{varepsilonf}
	\varepsilon_{abcd} = f(y) \: dy^0_a \wedge dy^1_b \wedge dy^2_c \wedge dy^3_d
\end{equation}
As indicated previously, a distinct $f$ would correspond to a different theory.

Now, let us consider solving the field equation \eqref{UGecs2}, ignoring for the moment the constraint. Suppose that we have found a solution given by $g_{ab} = g_{\mu \nu} (y) dy^\mu_a \otimes dy^\nu_b$, and the determinant $\det(g_{\mu \nu} (y))\equiv g(y)$ can be obtained. In this manner, we consider the volume element associated with the metric,
\begin{equation}\label{espilong}
	\epsilon^{(g)}_{abcd} = \sqrt{-g(y)} \: dy^0_a \wedge dy^1_b \wedge dy^2_c \wedge dy^3_d.
\end{equation}
Here, it is important to mention that (as emphasized for example in \cite{Waldbook}), the canonical way to characterize $g(y)$ is to claim that such a quantity connects the 4-volume element--corresponding to the metric--with the fiducial volume element in local coordinates $y^\mu$, namely the object  $dy^0_a \wedge dy^1_b \wedge dy^2_c \wedge dy^3_d$.  In addition, note that the integration of $\epsilon^{(g)}_{abcd}$ is defined in a purely geometric manner, regardless of the choice of the coordinate system, once the metric $g_{ab}$ is fixed.

Furthermore, after inspecting Eqs. \eqref{varepsilonf} and \eqref{espilong}, if it turned out that $\sqrt{-g(y)} = f(y)$, the task of finding the solutions of the UG field equations \eqref{UGecs2} and \eqref{constraint} would be complete.  In principle, this case might occur just by pure chance.
On the contrary, if it turned out that $\sqrt{-g(y)} \neq f(y)$, then, we would have to conclude that this is not a solution to our theory involving the originally given $f$.  That statement is correct. However, we should point out that this case could be considered as the solution to another theory. Specifically, the one which had a different $f$, say $\tilde f$, such that $\tilde{f} (y) = \sqrt{-g(y)}$.  So, the following question arises: how is this latter solution related to the solution of the original theory involving the given fiducial volume element described by $f(y)$?  We will address this question below.

A simple procedure is to regard the solution of \eqref{UGecs2} we have found, $g_{\mu \nu}  (y)$, as really being given in different coordinates. In other words, we can use the same functional form but changing the label of the coordinates, i.e. change the notation in the solution and rename the coordinates $x^\mu$.  Hence, we consider a new solution defined as $\texttt{g}_{ab} \equiv g_{\mu \nu} (x) dx^\mu_a \otimes dx^\nu_b$, where $g_{\mu \nu} (x)$ is obtained simply by substituting the label $y^\mu$ by the label $x^\mu$ in the arguments of the same functions. It is clear that the functions $g_{\mu \nu} (x)$ satisfy the same differential equations with respect to the coordinates $x^\mu$, as the original functions $g_{\mu \nu} (y)$ satisfy the equations with respect to the coordinates $y^\mu$ (as long as we also substitute in a similar way the functions characterizing the energy-momentum tensor of matter). This is a consequence of the covariance of \eqref{UGecs2}.

The new solution $\texttt{g}_{ab}$ can be expressed in the ``old'' coordinates (the exact relation between the two is yet to be specified) by setting $y^\mu = F^{(\mu)} (x)$, where $x$ represents collectively all the coordinates $x^\nu$. Evidently, the metric at any point $p$ can be expressed in either coordinates:
\begin{equation}\label{relacionmetricas}
	\texttt{g} _{ab} (p) \equiv g_{\mu \nu}(x(p)) \: dx^\mu_a \otimes dx^\nu_b = \tilde{g}_{\alpha \beta} (y(p)) \: dy^\alpha_a \otimes dy^\beta_b.
\end{equation}
As it is well known,  under a change of coordinates $\tilde g_{\alpha \beta} \frac{\partial y^\alpha}{\partial x^\mu}  \frac{\partial y^\beta}{\partial x^\nu} = g_{\mu \nu}$, and thus
\begin{equation}\label{relaciondets}
	\text{det} \: \tilde g_{\alpha \beta} = \left( \text{det}  \: \frac{\partial F^\mu }{\partial x^\alpha}  \right)^{-2} 	\text{det} \:  g_{\mu \nu}.
\end{equation}
We can now easily arrange to fix the coordinate transformation to be such that
\begin{equation}\label{transfposta}
	-\text{det} \: \tilde{g}_{\alpha \beta} (y) \equiv - \tilde{g} (y) = f(y)^2.
\end{equation}
In fact, there are actually many ways to do that, but it is easy to see that it can be achieved even by making just one of the functions $F^\mu$ to be nontrivial, while the other three are ``the identity''. We can, for instance, focus on the case where it is only $F^0$, and, moreover, we can simply set it to depend just on $x^0$ (this will be more clear in the example provided at the end of this section).

Therefore, after finding the suitable transformation satisfying Eq. \eqref{transfposta}, we would have that the 4-volume associated with the new solution $\texttt{g}_{ab}$ is:
\begin{eqnarray}\label{constraint2}
	\epsilon_{abcd}^{(\texttt{g})} &=& \sqrt{-\tilde g (y) } \: dy^0_a \wedge dy^1_b \wedge dy^2_c \wedge dy^3_d \nn
	&=& f(y) \: dy^0_a \wedge dy^1_b \wedge dy^2_c \wedge dy^3_d = \varepsilon_{abcd}.
\end{eqnarray}
Thus, the new metric $\texttt{g}_{ab}$  is now a solution to both the field equations \eqref{UGecs2} and the constraint \eqref{constraint}.

On the other hand, we might instead consider the full situation above, and describe everything in terms of the new coordinates $x^\mu$. In that case, the initially fiduciary volume element is
\begin{eqnarray}\label{varepsilonx}
	\varepsilon_{abcd} &=& f(y) \:  dy^0_a \wedge dy^1_b \wedge dy^2_c \wedge dy^3_d \nn
	&=& \sqrt{-\tilde g  } \: dy^0_a \wedge dy^1_b \wedge dy^2_c \wedge dy^3_d \nn
	&=&  \sqrt{ - g } \left( \text{det}  \: \frac{\partial F^\mu }{\partial x^\alpha}  \right)^{-1}   dy^0_a \wedge dy^1_b \wedge dy^2_c \wedge dy^3_d  \nn
	&=& \sqrt{-g(x)} \: dx^0_a \wedge dx^1_b \wedge dx^2_c \wedge dx^3_d,
\end{eqnarray}
where in the second and third lines we have used Eqs. \eqref{transfposta} and Eq. \eqref{relaciondets}, respectively, and in the final line we performed a transformation  between the coordinates $y^\mu$ and $x^\mu$ on the 4-form.

The last expression in Eq. \eqref{varepsilonx} indicates that the procedure we have employed is, in effect, equivalent, at the practical level, to simply pretend that we were from the start dealing with a theory where the $f$ that was initially given ``happened to coincide'' with what was needed in order to solve the constraint equation \eqref{constraint}. In other words, we could consider the volume element described by $f$ as representing the fiduciary volume element expressed in coordinates $y$, which differed from the coordinates in which we were analyzing the metric functions $g_{\alpha \beta} (x)$ in the first place.

Lastly, it is important to mention that, once we have found a solution to the field equations \eqref{EFE}, or equivalently the trace-free equations \eqref{UGecs2}, and chosen to  impose  the conservation of the energy-momentum tensor,  the physical solutions within UG are indistinguishable from GR.  That is to say, if one assumes $\nabla_a T^{ab}=0$, then, in practice UG is equivalent to GR. The only difference is that the cosmological constant is not naturally identified as the vacuum energy, but rather as an integration constant whose value, in principle, is arbitrary \cite{Anderson:1971pn, Henneaux:1989zc, UG2, Weinberg89}. On the other hand, when considering relaxing the requirement of conservation of energy-momentum, novel and often interesting possibilities appear \cite{Perez2019,sudarskyH0,sudarskyPRL2,Amadei2021,deCesare2021,Cuzinatto2022,Yang2022,Leon2022,Pia2023}.

\subsection{A cosmological example}

As a simple and concrete illustration of the previous discussion, we present an example that also illustrates one of the most common mistakes found in several discussions of UG.
We start by assuming that we have solved the standard EFE in the traditional cosmological setting, where the solution corresponds to a spatially flat Friedmann-Lema\^itre-Robertson-Walker (FLRW) spacetime. Using the coordinates $\{t,x,y,z\}$, the metric is represented by the line element
\begin{equation}\label{gFLRW}
	ds^2 = -dt^2 + a(t)^2 (dx^2 + dy^2 +dz^2).
\end{equation}
corresponding to, say, a radiation filled universe, so that $a(t) = C  t^{1/2}$. A common statement found in the literature regarding UG is: \emph{the only valid solutions of UG are such that the corresponding volume element must be of unitary determinant or $\sqrt{-g}=1$}, i.e. $f=1$ in our previous discussion. According to that, the metric characterized by Eq. \eqref{gFLRW} would not represent a solution of UG, because the determinant of the metric yields $\sqrt{-g} = a(t)^3 \neq 1$.

However, we can consider a {\it new solution} given in coordinates $\lbrace{\tau,  x,  y,  z }  \rbrace$ given by
\begin{equation}\label{gFLRW2}
	ds_{\textrm{New}}^2 = -d\tau^2 + {\tilde a}(\tau)^2 (dx^2 + dy^2 +dz^2).
\end{equation}
with ${\tilde a}(\tau) = C {\tau}^{1/2}$, which we must emphasize, should not be taken as corresponding to a standard change of coordinates, but rather to a new solution, in different coordinates. The point is that this line element corresponds to a radiation filled FLRW universe, which at {\it time} $\tau = t$, has the same density as that of Eq. (\ref{gFLRW}) at time $t$.
Let us now write this new solution in the original variables $\lbrace{t,  x,  y,  z }  \rbrace $ by fixing the relation between $t$ and $\tau $, so that ${\tilde a} (\tau)^{3} d \tau  = dt$. By setting $t(\tau=0) =0$, we have $t= ( 2/5)  C^3 \tau^{5/2}$ and ${\tilde a }(t) = [(5/2) C^2 t ]  ^{1/5}$. In the new coordinates, the line element \eqref{gFLRW} is thus,
\begin{equation}\label{gFLRWG}
	ds_{\textrm{New}}^2 = -{\tilde a}(t)^{-6} dt^2 + {\tilde a}(t)^2 (dx^2 + dy^2 +dz^2).
\end{equation}
It is clear that, in these coordinates, the new solution has determinant $\sqrt{-g}=1$.
However, it is still the {\it new} solution of Eq. \eqref{gFLRW2}. We could just as well have kept using the coordinates $\lbrace{\tau,  x,  y,  z } \rbrace$, and simply note that the fiduciary volume element $dt \wedge dx \wedge dy \wedge dz$, when expressed in the new coordinates, takes the form ${\tilde a} (\tau)^{3} d \tau \wedge dx \wedge dy \wedge dz$ which, in fact, coincides with the volume element of the metric in Eq. \eqref{gFLRW2}. Needless is to say that the name we give  the {\it time} coordinates is irrelevant, and we could, as well, have rewritten Eq. \eqref{gFLRW2} and the corresponding volume element simply making the replacement $\tau\to t$.

Thus, the ``problem'' was that we were not expressing the original metric in Eq. \eqref{gFLRW} in the appropriate coordinates (i.e. the ones in which the volume  element was simply the wedge product of the coordinate differential one forms). However, as we have seen, once that small detail is fixed, it is clear that the  original FLRW spacetime, with metric given by Eq. \eqref{gFLRW}, must be considered as a valid solution within UG. Alternatively, we could have said that the requirement $f=1$ alluded to the coordinates $\{t,x,y,z\}$ as considered in Eq. \eqref{gFLRWG}, while in the coordinates $\{\tau,x,y,z\}$ as in Eq. \eqref{gFLRW2} (i.e. characterizing the usual flat FLRW spacetime) the constraint was simply expressed by demanding $f = a(\tau)^3$.

The point is that, given a solution to Einstein's equations, such as Eq. \eqref{gFLRW}, the procedure described above consistently solves both the UG field equations \eqref{UGecs2} and the constraint \eqref{constraint}. Moreover, if one chooses to impose that $\nabla^a T_{ab}=0$, where  $T_{ab}$ might correspond to the energy-momentum tensor of a perfect fluid, then the cosmological model based on the flat FLRW spacetime within GR is identical to the one obtained in UG (after setting the cosmological constant, which, in that case, is just an integration constant to zero).

We note that the procedure we have used completely bypasses a standard objection that is raised when the approach is presented as  ``simply changing variables''. When that is done, it is often argued that {\it ``such approach overlooks the fact that a change of variables might change the determinant of the metric,  but it would also change the expression for the fiducial volume element. Thus, if those were different  when expressed in the first set of coordinates, they would also differ in the new set of coordinates''}. In the above, we have made it clear how the procedure must be viewed in order to be completely rigorous and strictly correct.

\section{Basic aspects of covariant perturbation theory}\label{Sec_perturb}

In this section, we will introduce a brief summary of covariant perturbation theory, so no original work is presented here. We will follow Refs.  \cite{adolfoRM,nakamura2006,nakamura2008,malik2008} closely. The motivation for this quick review is to present the subject in a complete geometrical manner.  For readers familiar with the topic, but who consider the formalism of covariant perturbation theory as ``infinitesimal coordinate transformations'' (e.g. as introduced in   \cite{mukhanov92,dodelson93,mukhanov2005}), we kindly suggest \emph{not to skip} this section.

\subsection{Fundamentals}\label{Sec_perturbA}

In perturbation theory, one deals with expressions of the type $Q \simeq Q_0 + \lambda \delta Q$, where $Q$ is a tensor field characterizing a physical quantity. Generally, the object $Q_0$ represents an exact solution to an equation corresponding to a known problem, $\delta  Q$ is a deviation from such solution, and the parameter $\lambda$ quantifies the smallness of the deviation from the original solution. However, in a covariant theory, e.g.  GR, that kind of expression is problematic because, by itself,  it is generically not defined in a precise manner. The tensor field $Q$ is evaluated at some point $\tilde p$ of the physical spacetime $(\mM,g_{ab})$, where $\mM$ denotes the spacetime manifold with a spacetime metric $g_{ab}$.  Meanwhile, $Q_0$ and $\delta Q$ are evaluated at a point $p$ of a ``background'' spacetime $(\mM_0,\overline{g}_{ab})$.
One would like to somehow ``identify'' $\tilde p$ and $p$,  but there is, in general, no canonical way to do  this. The background spacetime is distinct from the physical spacetime $(\mM,g_{ab})$. The former was introduced only to perform the perturbative analysis, while the latter is presumably describing the actual physical situation of interest. It is worth keeping in mind that the choice of the coordinate charts for each spacetime are not directly comparable, and, in fact, the introduction of coordinates too early in the discussion often generates more confusion. Therefore, an important aspect of covariant perturbation theory is to characterize in a precise way the meaning of the expression
\begin{equation}\label{eqQ}
	\overbrace{Q(\tilde p )}^{\in (\mM, g_{ab}) } \simeq \overbrace{Q_0 (p) + \lambda \delta Q(p)}^{   \in (\mM_0, \overline{g}_{ab})  }.
\end{equation}

The perturbative analysis could be concerned with the metric itself, i.e. $Q$ might correspond to the spacetime metric $g_{ab}$. For instance, we can consider again the situation where we have two different physical spacetimes with some matter fields defined on them, and  we want to compare $(\mM_0,\overline{g}_{ab}, \overline{\Psi}_M)$ with $(\mM, g_{ab}, \Psi_M)$. In order to do that, we introduce a diffeomorphism $\Phi: \mM_0 \mapsto \mM$, (we will be assuming that the two differential manifolds are diffeomorphic). Then, the differences
\begin{equation}\label{diffs}
	\delta g_{ab} \equiv \Phi^* g_{ab} - \overline{g}_{ab}, \qquad \delta \Psi_M  \equiv  \Phi^* \Psi_M -  \overline{\Psi}_M,
\end{equation}
are well defined operations on $\mM_0$, where $\Phi^*$ is the \textit{pullback} associated to $\Phi$.

The differences $\delta g_{ab} $ and $ \delta \Psi_M  $ are what is often considered as \textit{perturbations} of the ``background'' spacetime $(\mM_0,\overline{g}_{ab}, \overline{\Psi}_M)$, which, for example, could be a spatially homogeneous and isotropic spacetime. On the other hand, $(\mM, g_{ab}, \Psi_M)$ might represent an inhomogeneous and anisotropic spacetime characterizing, say, our universe. Hence, in that case, the perturbations describe a small deviation from the homogeneity and isotropy that is encoded in the background spacetime.

Evidently, the expression for the differences \eqref{diffs} depends on the choice of $\Phi$. The selection of a particular $\Phi$ is called the \textit{gauge choice} in perturbation theory, and given the covariant nature of the formulation, the choice of such a diffeomorphism between $\mM_0$ and $\mM$ is far from unique. This is \textit{not} an issue of coordinates, but it can be confused with one because, intuitively, one associates ``general covariance'' with the generic prescription about the equivalence of all  coordinate's systems used in working with the theory. The degree of freedom reflected in the choice of $\Phi$ is called the \textit{gauge degree of freedom}, and it does not represent a \emph{physical} degree of freedom. There are two options to deal with the issue of gauge choice:
\begin{enumerate}
	\item Fix the gauge, and proceed with the calculations in that chosen gauge.
	
	\item Work with \textit{gauge invariant} quantities, i.e. look for combinations that are invariant under ``small changes of $\Phi$'' \footnote {Note, however, that, in this case, a choice of primary diffeomorphism, with respect to which one might consider the small changes, should have been made.}.
\end{enumerate}

Both approaches have their advantages and disadvantages, depending on the particular situation under consideration. There is a substantial amount of work in the literature covering the two methods (see the review \cite{malik2008}). Nevertheless, it is clear that physical observables are independent of  either option. In the next subsection, we will be more precise by introducing the formalism needed for all the main elements involved in the two aforementioned approaches.

\subsection{The formalism}

In order to ensure transparency of the description, it is convenient to consider perturbation theory in terms of a one-parameter family of diffeomorphisms, and, in fact, making the whole set up using a single higher dimensional manifold. That is, we begin by considering a manifold $\mN = \mM \times  \mathbb{R}$ with dim $\mN =$ dim $\mM + 1$, so for a 3+1 dimensional spacetime, dim $\mN=5$. We denote the parameter labeling the elements of the group of diffeomorphism by $\lambda$, and use it to perform the perturbative expansion. We take $\mM_0 := \mN|_{\lambda=0}$ and $\mM \equiv \mM_{\lambda} :=  \mN|_{\lambda }$ for some $\lambda \in \mathbb{R}$ satisfying $0< \lambda \ll 1$ (which, for convenience, we will not take as fixed in the analysis, but which is understood to take a finite value in actual applications).

A point in $\mN$ is denoted by the pair $(p,\lambda)$, where $p \in \mM_{\lambda}$ and a point in $\mM_0$ corresponds to $(p,0)$. Thus the  1-parameter group of diffeomorphisms allows us to identify points between the various hypersurfaces $\mM_\lambda$ in $\mN$. In particular, we can take the map $\chi_\lambda : \mN \mapsto \mN$, and focus our attention on its restriction to $\mM_0 $, namely $\chi_\lambda : \mM_0 \mapsto \mM_\lambda$.

Given the aforementioned construction, we can consider the perturbations of the  tensor field $Q$ as comparisons between $Q \in \mM_\lambda$ and $Q_0 \in \mM_0$ through the map $\chi_\lambda^*$ [in the same sense as the differences \eqref{diffs}].  The selection of a particular $\chi_\lambda$ corresponds to the\textit{ gauge choice} that we have discussed in the previous subsection, i.e. $\chi_\lambda$ plays the same role as the map $\Phi$ introduced there. Furthermore, the 1-parameter diffeomorphism $\chi_\lambda$ satisfies the following properties:
\begin{equation}\label{key}
	\chi_{\lambda_1+\lambda_2} = \chi_{\lambda_1} \circ \chi_{\lambda_2} = \chi_{\lambda_2} \circ \chi_{\lambda_1}, \qquad \chi_0 = \mathbb{I}.
\end{equation}

In order to define, in a precise manner, the perturbative expansion at order $k$ of the tensor field $Q$, we consider that the 1-parameter group of diffeomorphisms is generated by the vector field $\xi^a_\lambda$ defined through the Lie derivative
\begin{equation}\label{lieQ}
	\pounds_\xi Q \equiv \lim_{\lambda \to 0} \frac{\chi_\lambda^* Q - Q }{\lambda},
\end{equation}
where $\chi_\lambda^* Q $ represents the \textit{pullback} of $Q$ (defined in $\mN$) by $\chi$.  The pullback  $\chi_\lambda^* Q $ maps the tensor field $Q$ in $\mM_\lambda$ to a tensor  $\chi_\lambda^* Q $ in $\mM_0$.

In this manner, $\chi_\lambda^* Q$ can be expanded as a Taylor series \cite{Bruni1996}
\begin{equation}\label{pullQtaylor}
	\chi_\lambda^* Q |_{\mM_0} = \sum_{k=0}^{\infty} \frac{\lambda^k}{k!} \pounds_{\xi_\lambda}^k Q|_{\mM_0}.
\end{equation}
Hence, each term of the expansion for $k \geq 1$ will correspond to the perturbation of the ``background value'' of  the physical variable $Q$ at order $k$. In order to be more explicit, we will focus on the first order perturbation. The expansion \eqref{pullQtaylor} is therefore explicitly expressed as
\begin{equation}\label{pullQorden1}
	\chi_\lambda^* Q(p)	 = Q_0(p) + \lambda \pounds_{\xi_\lambda} Q|_{\mM_0} (p) + \mathcal{O}(\lambda^2),
\end{equation}
where $p \in \mM_0$, and $Q_0 = Q |_{\mM_0}$, i.e. the background value of $Q$. Consequently, with the gauge choice $\chi_\lambda$ the difference
\begin{equation}\label{diffQ}
	\Delta^\chi Q_\lambda \equiv \chi_\lambda^* Q(p) - Q_0(p)
\end{equation}
is well defined in $\mM_0$.

Substituting \eqref{pullQorden1} in \eqref{diffQ} defines the first order perturbation of $Q$ with the gauge choice $\chi_\lambda$ as
\begin{equation}\label{pertQorden1}
	\delta_\chi Q \equiv   \lambda \pounds_{\xi} Q|_{\mM_0}.
\end{equation}

Equations \eqref{diffQ} and \eqref{pertQorden1}, can be used to characterize, in a precise manner,  what it means to consider ``small perturbations'' of the background tensor field $Q_0$. In other words, at first order in Eq. \eqref{pullQorden1}, the tensor $\chi_\lambda^* Q$, which is defined in $\mM_0$, is considered as an approximation (at first order) to the physical tensor field $Q$, which is defined in $\mM_{\lambda}$, with the gauge choice given by $\chi_\lambda$.

\begin{figure}[t!]
	\centering
	\includegraphics[scale=0.5]{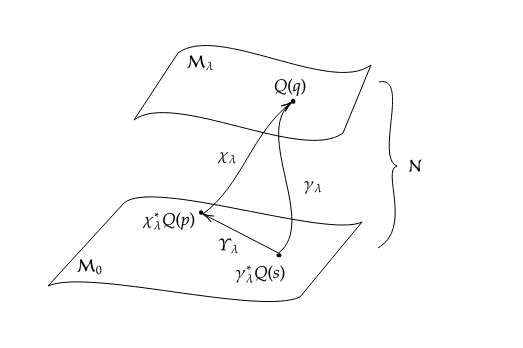}
	\caption{For a fixed point $q$ on $\mM_\lambda$, we can choose a gauge $\chi_\lambda$ or $\gamma_\lambda$, in order to identify $q$ with $p \in \mM_0 $ or with $s \in \mM_0 $.  The map $\Upsilon_\lambda$, which maps the point $s$ to the point $p$ is then formed by  $\Upsilon_\lambda = \chi_\lambda^{-1} \circ \gamma_\lambda$.   }
	\label{fig_mapeo1}
\end{figure}

Let us now focus on the issue of \textit{gauge transformation}. To this end, we assume that there are two possible gauge choices $\chi_\lambda$ and $\gamma_\lambda$ with the generating vector fields $X^a$ and $Y^a$, respectively. If $X^a$ and $Y^a$ have different tangential components to each $\mM_\lambda$, then we refer to them as \textit{different gauge choices}. Moreover, a \textit{gauge transformation} is regarded as the change of the gauge choice between $\chi_\lambda$ and $\gamma_\lambda$, which is given by the diffeomorphism
\begin{equation}\label{difftransfgauge}
	\Upsilon_\lambda\equiv (\chi_\lambda)^{-1} \circ  \gamma_\lambda : \mM_0 \mapsto \mM_0.
\end{equation}
The diffeomorphism $\Upsilon_\lambda$ is a map for each value of $\lambda \in \mathbb{R}$, and reflects the difference in the identification among points of  $\mM_0$ with those in $\mM_\lambda$ that results from the two gauge choices.
Furthermore, $\Upsilon_\lambda$ induces a pullback from the representation $^{\chi}Q_\lambda$   in the gauge choice $\chi_\lambda$ to the representation $^{\gamma}Q_\lambda$  in the gauge choice $\gamma_\lambda$, see Fig. \ref{fig_mapeo1}. The pullback corresponding to $\Upsilon_\lambda$ acts as
\begin{equation}\label{pullbackupsilon}
	\:^{\gamma}Q_\lambda \equiv \gamma_\lambda^*Q|_{M_0} = (\chi_\lambda^{-1} \gamma_\lambda)^*(\chi_\lambda^* Q)|_{\mM_0} = \Upsilon_\lambda^* \:^{\chi}Q_\lambda.
\end{equation}

The object $^\gamma Q_\lambda=\Upsilon_\lambda^* \:^{\chi}Q_\lambda$ can be expressed as a ``Taylor series'',
\begin{eqnarray}\label{taylorupsilon}
	^{\gamma} Q_\lambda  &=&  \Upsilon_\lambda^* \:^{\chi}Q_\lambda  \nn
	&=& \:^\chi Q + \lambda \pounds_{\xi_1}  \:^\chi Q +  \frac{\lambda^2}{2} \{   \pounds_{\xi_2} + \pounds_{\xi_1}^2 \}  \:^\chi Q + \mathcal{O} (\lambda^3) , \nn
\end{eqnarray}
where $\xi_{1,2}^a$ are the first two generators of $\Upsilon_\lambda$, and 	$\:^{\chi}Q  \equiv \chi_\lambda^*Q|_{M_0}$. Also note that in Eq. \eqref{taylorupsilon}, the expansion corresponding to the pullback associated to $\Upsilon_\lambda$ is not  exactly a Taylor series. The reason is that $X^a$ and $Y^a$ do not commute,  in general.

The detailed procedure to obtain Eq. \eqref{taylorupsilon} is shown in Refs. \cite{adolfoRM,nakamura2006,Sonego1997}. However, at first order, Eq. \eqref{taylorupsilon} coincides exactly with the first order term of the corresponding Taylor series.  We will continue the analysis at first order only.

From Eqs. \eqref{taylorupsilon} and \eqref{pullQorden1}, then
\begin{equation}\label{gaugetransf}
	\delta_\chi Q - \delta_\gamma Q = \pounds_{\xi_1} Q_0.
\end{equation}
Additionally, it can be shown \cite{Bruni1996,nakamura2006,adolfoRM}  that the generator $\xi_1^a$, in terms of the generators $X^a$ and $Y^a$, is given as
\begin{equation}\label{xi1}
	\xi_1^a = Y^a - X^a.
\end{equation}
Thus, Eq. \eqref{gaugetransf}, together with Eq. \eqref{xi1}, defines the gauge transformation between the two possible gauge choices: $\chi_{\lambda}$ and $\gamma_\lambda$.

Finally, for the sake of completeness, we say that a tensor field $Q$  in $\mN$ is  \textit{totally gauge invariant} if $\:^{\chi}Q_\lambda = \:^{\gamma}Q_\lambda $  for any pair of gauge choices $\chi$ and $\gamma$, so $\delta^{(k)}_\chi Q = \delta^{(k)}_\gamma Q $ at any order $k$.  A more relaxed approach is to define \textit{gauge invariant at order $n$} if and only if, for any two gauges $\chi$ and $\gamma$,
\begin{equation}\label{giorden1}
	\delta^{(k)}_\chi Q = \delta^{(k)}_\gamma Q  \qquad \forall k, \qquad \textrm{with} \qquad k<n .
\end{equation}

The previous definition is used in Ref. \cite{Bruni1996} to show that a tensor field $Q$ is gauge invariant at order $n \geq 1$, if and only if, $\pounds_\xi \delta^{(k)} Q =0$, for any vector field $\xi^a$ defined in $\mM_0$, and $ \forall k < n$. Therefore, one has a generalization of the Stewart-Walker's Lemma \cite{Walker1974}: The perturbations of a tensor field $Q$ at order $n$ are gauge invariant, if and only if, $Q_0$ and all perturbations at order $< n$ are trivial in any gauge, i.e. one of the following holds: the perturbations are zero, constants, or linear combinations of products of Kronecker deltas.

\section{The perturbed FLRW spacetime in unimodular gravity}\label{Sec_UGpert}

We are now interested  in applying the previous formalism to the FLRW spacetime. In this way, we will address the cosmological perturbation theory in UG. Our motivation is to show that, contrary to widespread knowledge, there are no differences or restrictions with respect to standard GR, when implementing the perturbation theory in UG.

\subsection{The perturbed FLRW spacetime}

For the background metric, we consider a spatially flat FLRW spacetime. In comoving coordinates with conformal time $\eta$, the background metric is given as $\overline{g}_{\mu \nu} = a^2(\eta) \eta_{\mu \nu}$, where $\eta_{\mu \nu}$ is the Minkowski's metric. Therefore, the background spacetime is perfectly homogeneous and isotropic in space. The components of the perturbed part of the metric, at first order, can be written as
\begin{equation}\label{defperturbgmunu}
	\delta g_{\mu \nu} \equiv h_{\mu \nu} = a^2 (\eta) \theta_{\mu \nu} (\x,\eta).
\end{equation}
The perturbed line element is then
\begin{eqnarray}\label{ds2}
	ds^2 &=& a^2 (\eta) [ -(1+\theta_{00}) d\eta^2 + 2 \theta_{0i} d\eta dx^i \nn
	&+& (\delta_{ij} + \theta_{ij}) dx^i dx^j ]
\end{eqnarray}
with $i,j=1,2,3$.

Taking into account that the hypersurfaces $\eta =$ constant, in the background metric, are maximally symmetric, the components $\{\theta_{00}, \theta_{0i}, \theta_{ij} \}$, associated to the first order metric perturbation $h_{ab}$ [see Eq. \eqref{defperturbgmunu}], can be further decomposed into three sets of variables:
\begin{enumerate} \item the \textit{scalar} parts $\{\psi,\phi,E,B\}$,  \item the \textit{vector} parts $\{ S_i,F_i\}$, which are divergence free i.e. $\partial_k S^k = 0 = \partial_k F^k$, and \item  the \textit{tensor} part $\mH_{ij}$, which is transverse and traceless, i.e. $ \partial_k \mH^{ ik} =0$, $\mH^{k}_{\: k} =0$.
\end{enumerate}
The names scalar, vector and tensor come from the transformation properties of such variables under spatial rotations in the  background spacetime. Hence, those variables are not truly scalar, vector or tensor fields in the geometrical sense; however, since the names are very widespread in the literature,  we will continue to use them in this work. In terms of the scalar, vector and tensor variables (also known as \textit{modes}), the components of the perturbed metric, at first order,  are traditionally expressed as \cite{malik2008}
\begin{subequations}\label{perturbgmunu}
	\begin{equation}\label{g00}
		\overline{g}_{00} + h_{00} = -a^2 (1+2 \phi)
	\end{equation}	
	\begin{equation}\label{g0i}
		\overline{g}_{0i} + h_{0i} = a^2 (\partial_i B - S_i )
	\end{equation}
	\begin{equation}\label{gij}
		\overline{g}_{ij} + h_{ij} = a^2[ (1 -2 \psi) \delta_{ij}  + 2 (\partial_{ij} E + \partial_{(i} F_{j)} ) +  \mH_{ij} ]
	\end{equation}
\end{subequations}
Note that due to the constraints on the vector and tensor modes (i.e. the divergence free and the transverse/traceless conditions), there are 10 total degrees of freedom left, but only 6 of them are physical, the same as any spacetime metric (in 4 dimensions).

\subsection{FLRW spacetime: perturbations and gauge transformations}

The decomposition made in Eqs. \eqref{perturbgmunu} is not unique. In particular, we can view Eqs. \eqref{perturbgmunu} as the components of the perturbed metric in a specific  gauge $\gamma$. This is, there is a physical spacetime $\mM$ which is not homogeneous and isotropic, with metric $g_{ab}$, and we will approximate it by employing the decomposition in Eqs. \eqref{perturbgmunu}, which are evaluated at the background spacetime $\mM_0$. More precisely, using Eqs. \eqref{pullQorden1} and \eqref{pertQorden1} for the metric $g_{ab}$ in $\mM$, we obtain
\begin{equation}\label{pertmert}
	\gamma^* g_{ab} = \overline{g}_{ab} + \delta_{\gamma} g_{ab},
\end{equation}
where the full equation is evaluated in the background spacetime, and we identify $\delta_{\gamma} g_{ab} = \:_{\gamma} h_{ab}$ with the components of $\:_{\gamma} h_{ab}$, as given in Eqs. \eqref{perturbgmunu}.  However, we are free to choose another gauge $\chi$, in which the values of any of the scalar, vector and tensor modes of the perturbed metric differ from the original ones (in the gauge $\gamma$), i.e. $\delta_{\chi} g_{ab} \neq \delta_{\gamma} g_{ab} $. Equation \eqref{gaugetransf} yields the gauge transformation (between the gauges $\chi$ and $\gamma$) of the metric perturbations,
\begin{equation}\label{gaugetransfhab}
	\:_{\chi} h_{ab} - \:_{\gamma} h_{ab} = \pounds_\xi \overline{g}_{ab} = 2 \nabla_{(a} \xi_{b)},
\end{equation}
where $\xi^a$ is a vector field in the background spacetime $\mM_0$, and is a generator of the gauge transformation. That vector can be decomposed as
\begin{equation}\label{defxi}
	\xi^a = \alpha (\partial_\eta)^a + \beta^i (\partial_{i})^a,
\end{equation}
with the spatial part being able to be split as
\begin{equation}\label{defbetai}
	\beta^i = \partial^i \beta + B^i,
\end{equation}
where the vector $B^i$ is divergence free $\partial_k B^k=0$. Therefore, from Eq. \eqref{gaugetransfhab} one can obtain explicitly the gauge transformation of the first order metric perturbations. For the scalar modes these are:
\begin{subequations}\label{scalartransf}
	\begin{equation}\label{phitransf}
		\:_{\chi} \phi = \:_{\gamma} \phi + \mH \alpha + \alpha' ,
	\end{equation}
	\begin{equation}\label{psitransf}
		\:_{\chi} \psi = \:_{\gamma} \psi - \mH \alpha ,
	\end{equation}
	\begin{equation}\label{Btransf}
		\:_{\chi} B = \:_{\gamma} B - \alpha + \beta' ,
	\end{equation}
	\begin{equation}\label{Etransf}
		\:_{\chi} E =\:_{\gamma} E + \beta ,
	\end{equation}
\end{subequations}
where $\mH \equiv a'/a$, and $'$ denotes derivative with respect to conformal time $\eta$. For the vector perturbations one obtains:
\begin{subequations}\label{vectortransf}
	\begin{equation}\label{Sitransf}
		\:_{\chi} S^i = \:_{\gamma} S^i -  B^{i} \:' ,
	\end{equation}
	\begin{equation}\label{Ftransf}
		\:_{\chi} F^i = \:_{\gamma}  F^i + B^{i} \:'.
	\end{equation}
\end{subequations}
Note that the scalar and vector modes \textit{are not} gauge invariant.  On the other hand, the first order tensor perturbation is found to be gauge invariant
\begin{equation}\label{hTTtransf}
	\:_{\chi} \mH_{ij} = \:_{\gamma}  \mH_{ij}.
\end{equation}

\subsection{The perturbed volume element: fixing the gauge}

As we have argued in Sect. \ref{Sec_perturbA}, the issue of the lack of a unique gauge choice is a consequence of the covariant nature of the theory, and the  methods (1) and (2) described there are two possible ways to address the issue. The point, however, is that both schemes are valid, and at the end, both must yield the same theoretical predictions for the physical observables.

In order to clarify some common misconceptions regarding the issue of gauge choice in cosmological perturbation theory within UG, we will continue by choosing option (1).  In addition, we will focus on the volume element, which is an important tensor field in UG. In principle, we have two volume elements in UG: the fiduciary volume element $\varepsilon_{abcd}$, and the one associated with the metric $\epsilon_{abcd}^{(g)}$ [see Eq. \eqref{constraint0}].  We now proceed to fix the gauge, i.e. we choose the gauge $\gamma$ characterized by Eqs. \eqref{perturbgmunu}, so we have
\begin{equation}\label{epsilonfpert}
	\gamma^* \varepsilon_{abcd} =\overline{\varepsilon}_{abcd}  + \delta_\gamma \varepsilon_{abcd},
\end{equation}
and
\begin{equation}\label{epsilongpert}
	\gamma^* \epsilon_{abcd}^{(g)} =  \overline{\epsilon}_{abcd}^{(g)} + \delta_\gamma  \epsilon_{abcd}^{(g)}.
\end{equation}

Let us concentrate on the volume element associated to the metric $\epsilon_{abcd}^{(g)}$. For the particular gauge $\gamma$, we can express $\gamma^* \epsilon_{abcd}^{(g)}$ as
\begin{eqnarray}\label{epsilongpert1}
	\overline{\epsilon}_{abcd}^{(g)} + \delta_\gamma  \epsilon_{abcd}^{(g)} &=& a^4  (1 + \phi - 3 \psi + \nabla^2 E) \nn
	&\times& d\eta_a \wedge dx^1_b \wedge dx^2_c \wedge dx^3_d,
\end{eqnarray}
where $\nabla^2 \equiv \partial_i \partial^i$.

With Eq. \eqref{epsilongpert1} at hand, we can identify one of the most common mistakes found in the literature when dealing with cosmological perturbation theory in UG. For instance, in Refs. \cite{Gao2014, Basak2015}, one is told that because of the ``UG constraint'' $g^{\mu \nu} \delta g_{\mu \nu} = 0$ (which comes from $\delta \sqrt{-g} =0$), and using the components $g_{\mu \nu}$ and $ \delta g_{\mu \nu}$ given by Eqs. \eqref{perturbgmunu}, the gauge freedom is limited
just to choices which ensure that
\begin{equation}\label{constraintbranden}
	\phi - 3 \psi + \nabla^2 E = 0.
\end{equation}

However, we see from Eqs. \eqref{epsilongpert} and \eqref{epsilongpert1} that by (erroneously) declaring as acceptable just the gauge choices that satisfy \eqref{constraintbranden}, is tantamount to saying that, in UG, we must restrict ourselves to choose a gauge, say $\gamma_\diamond$, such that $\gamma^*_{\diamond} \epsilon_{abcd}^{(g)} = \overline{\epsilon}_{abcd}^{(g)}$, or equivalently
\begin{equation}\label{constraintbranden99}
	\:_{\gamma_\diamond}\delta \epsilon_{abcd}^{(g)} =0.
\end{equation}
While one can certainly {\it choose} to work with such gauge choices, as we have seen, there is nothing in the theory which forces us to do so. Thus, imposing  Eq. \eqref{constraintbranden} for the cosmological perturbation theory, as a consequence of the ``UG constraint'' $g^{\mu \nu} \delta g_{\mu \nu} = 0$, is simply incorrect.

In fact, as we have analyzed (and justified) in Sect. \ref{Sec_UGsolA}, in UG one can always proceed as if the given fiduciary volume element coincides exactly with the volume element associated to the metric. In particular, we can always ensure that
\begin{equation}\label{constraintpert}
	\gamma^* \epsilon_{abcd}^{(g)}|_{\mM_0} = 	\gamma^* \varepsilon_{abcd}|_{\mM_0}.
\end{equation}
However, note that the previous condition is not \eqref{constraintbranden}, but simply:
\begin{equation}\label{constraintpert2}
	\overline{\epsilon}_{abcd}^{(g)} = \overline{\varepsilon}_{abcd} \qquad  \:_\gamma \delta \epsilon_{abcd}^{(g)} 	= \:_\gamma \delta \varepsilon_{abcd},
\end{equation}
which does not constrain the modes $\phi$, $\psi$, and $E$ in any way.
Equation \eqref{constraintpert} [or Eq. \eqref{constraintpert2}] satisfies the correct UG constraint, which we have found in Eq. \eqref{constraint0}. It is important to mention that Eq. \eqref{constraintpert} is valid in any generic gauge. For the particular gauge $\gamma$, we have $\gamma^* \epsilon_{abcd}^{(g)}$ as in Eq. \eqref{epsilongpert1}.  Thus, there is no restriction of any kind preventing the use of Eq. \eqref{constraintpert} in the gauge $\gamma$, and there is no constraint imposing the use of a gauge such that Eq. \eqref{constraintbranden} is satisfied.

\subsection{Gauge transformation of the volume element: The Newtonian gauge}\label{Sec54}

To further analyze the error of enforcing Eq. \eqref{constraintbranden} as the ``UG constraint'', we will compute the gauge transformation of  $\:_\gamma \delta \epsilon_{abcd}^{(g)}$ to another specific gauge, say $\:_\chi \delta \epsilon_{abcd}^{(g)}$, and then focus on the so called Newtonian gauge as an example.

We start by choosing $\xi^a$, which was introduced in Eq. \eqref{defxi}, as the generator of the gauge transformation between gauges $\chi$ and $\gamma$.  Therefore, according to Eq. \eqref{gaugetransf}, the gauge transformation for the perturbation of the metric volume element is,
\begin{eqnarray}\label{epsilongaugetransform}
	\:_\chi \delta \epsilon_{abcd}^{(g)} 	- \:_\gamma \delta \epsilon_{abcd}^{(g)}  &=& a^4  ( \alpha' + 4 \mH \alpha  +   \nabla^2 \beta)  \nn
	&\times& d\eta_a \wedge dx^1_b \wedge dx^2_c \wedge dx^3_d.
\end{eqnarray}

Here, we note that if one were to impose the condition $g^{\mu \nu} \delta g_{\mu \nu} = 0$ (erroneously referred as ``the UG constraint''), then one would limit the  consideration of gauge transformations to those that are compatible with \eqref{constraintbranden}. That is, according to such a misunderstanding, if one were to  consider another gauge choice $\chi_\diamond$, one would have to demand that $\:_{\chi_\diamond}\delta \epsilon_{abcd}^{(g)} = 0 =  \:_{\gamma_\diamond}\delta \epsilon_{abcd}^{(g)}$. From Eq. \eqref{epsilongaugetransform}, we see that such requirement is equivalent to demanding that the scalar components of the generator $\xi^a$ satisfy:
\begin{equation}\label{constraintbranden2}
	\alpha' + 4 \mH \alpha  +   \nabla^2 \beta = 0.
\end{equation}
However, once again, we emphasize that there is no special requirement in the adequate UG constraint Eq. \eqref{constraintpert} [or \eqref{constraintpert2}],  that  would restrict the generators of the gauge transformation to be such so as to enforce Eq. \eqref{constraintbranden2}.

As a concrete example, we can consider the Newtonian (also known as longitudinal) gauge. This gauge is characterized by setting
\begin{equation}\label{EBnewt}
	E_{\chi_N} = B_{\chi_N} = 0
\end{equation}
in Eqs. \eqref{perturbgmunu}, where ${\chi_N}$ denotes that we are working in the Newtonian (longitudinal) gauge. On the other hand, we can take as given the metric perturbations in the gauge $\gamma$. Therefore, the scalar components of the generator of the gauge transformation between the Newtonian ${\chi_N}$ and $\gamma$ gauges can be found directly from Eqs. \eqref{Etransf} \eqref{Btransf} by fixing $E_{\chi_N} = B_{\chi_N} = 0$ on the left-hand-side of such equations, this is
\begin{equation}\label{scalarcomptransf}
	\alpha = B_\gamma-E_\gamma', \qquad \beta = -E_\gamma.
\end{equation}
Equation \eqref{scalarcomptransf} allow us to obtain the non-vanishing scalar perturbations from Eqs. \eqref{phitransf}, \eqref{psitransf}, these are
\begin{subequations}\label{transfscalN}
	\begin{equation}
		\phi_{\chi_N} = \phi_\gamma + \mH (B_\gamma-E_\gamma') + (B_\gamma'-E_\gamma''),
	\end{equation}
	\begin{equation}
		\psi_{\chi_N} = \psi_\gamma - \mH (B_\gamma-E_\gamma').
	\end{equation}
\end{subequations}

Additionally, by substituting Eq. \eqref{scalarcomptransf} in Eq. \eqref{epsilongaugetransform}, we can obtain the gauge transformation between $\chi_N$ and $\gamma$ for the perturbation of the metric volume element,
\begin{eqnarray}\label{epsilongaugetransformN}
	\:_{\chi_N} \delta \epsilon_{abcd}^{(g)} 	- \:_\gamma \delta \epsilon_{abcd}^{(g)} & =&  a^4  [ B_\gamma'-E_\gamma'' \nn
	& +&  4 \mH (B_\gamma-E_\gamma')  -   \nabla^2 E_\gamma ] \nn
	&\times& d\eta_a \wedge dx^1_b \wedge dx^2_c \wedge dx^3_d.
\end{eqnarray}
Finally, substituting Eq. \eqref{transfscalN} in Eq. \eqref{epsilongaugetransformN}, and also using Eq. \eqref{epsilongpert1}, we can find the expression for the metric volume element in the Newtonian gauge. Therefore,
\begin{eqnarray}\label{epsilongpertN}
	\overline{\epsilon}_{abcd}^{(g)} + \:_{\chi_N} \delta \epsilon_{abcd}^{(g)} &=& a^4  (1 +\phi_{\chi_N} - 3 \psi_{\chi_N} ) \nn
	&\times& d\eta_a \wedge dx^1_b \wedge dx^2_c \wedge dx^3_d.
\end{eqnarray}

\begin{figure}[t!]
	\centering
	\includegraphics[scale=0.5]{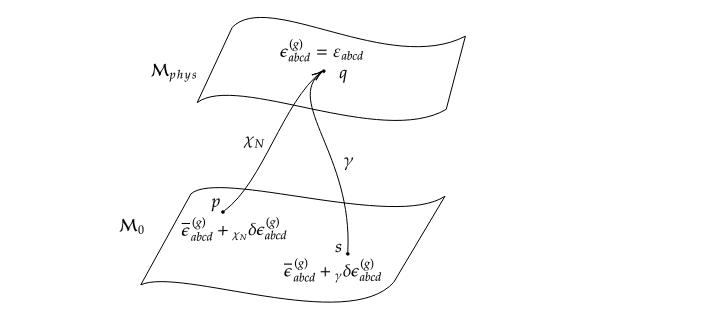}
	\caption{We have shown that the UG constraint $\epsilon_{abcd}^{(g)} = \varepsilon_{abcd}$ is trivialized both in $\mM_{\textrm{phys}}$ and $\mM_0$. Consequently, there are no restrictions on how to choose any of the gauges, $\chi_N$ or $\gamma$.}
	\label{fig_mapeo2}
\end{figure}

At this point, we note that if we were to follow the standard literature on the subject, e.g. Refs. \cite{Gao2014, Basak2015}, and erroneously impose the condition mistakenly described as the ``UG constraint'', namely, $g^{\mu \nu} \delta g_{\mu \nu} = 0$, we would have been prevented from working with the Newtonian gauge. That is, by unnecessarily setting $E_{\chi_N}=B_{\chi_N}=0$,  (i.e. the Newtonian gauge), one would automatically fix $\alpha$ and $\beta$ as in Eq. \eqref{scalarcomptransf}, which generically fails to satisfy the so called ``constraint condition'' \eqref{constraintbranden2}.
However, as we have shown previously, the analysis behind such aspect of the standard literature is simply flawed, and there is no restriction in UG that prevents us to choose the Newtonian gauge or any other gauge.
In other words, when the appropriate UG constraints \eqref{constraint0} are taken into account, which, in fact, are trivialized both in the physical and background spacetime (as shown in Sec. \ref{Sec_UGsolA}), there are no limitations on the relationship between the volume element(s) defined in $\mM_{\textrm{phys}}$ and $\mM_0$, see Fig \ref{fig_mapeo2}.

\section{Conclusions}\label{conclusions}

In this work, we have reviewed the general notion of diffeomorphism invariance and offered arguments to address some of the confusions that arise in discussions on the subject. In particular, we have clarified the difference between the \emph{tautological} nature of diffeomorphism invariance (as it applies to all theories specified by action functionals that are expressed as integrals over manifolds), and the more nuanced notion of \emph{dynamical} diffeomorphism invariance, which is the notion that is often used in Physics. We have expanded on that discussion as it concerns theories that are said to be less than fully diffeomorphism invariant, and, in particular, those that are said to be invariant under a subclass of diffeomorphisms.

Specifically, we have focused on the theory known as Unimodular Gravity (UG), which is often described as being invariant {\it only} under the {\it volume preserving diffeomorphisms}. When UG is presented using this perspective, it is often argued that the potential for non-conservation of the energy-momentum tensor arises in a close connection with that feature, as illustrated in the derivation of Eq. \eqref{conservTab} in Section \ref{Sec21}. However, it is important to note that any theory derived from an action, expressed as a well-defined integral over a manifold, is, by its very construction, invariant under a \emph{general} one-parameter family of diffeomorphisms when variations are performed on all geometric elements of the action, whether dynamical or not. Thus, the possible non-conservation of the energy-momentum tensor, as described in Eq. \eqref{conservTab}, can be equally well obtained, without restricting consideration just to the volume-preserving diffeomorphisms as shown in Section \ref{Sec22}.

We have clarified what does it mean, in practice, to find a solution to the trace-free part of Einstein's field equations (i.e. the UG equations for the gravitational field), and we have argued that the often used {\it auxiliary requirement} that $\sqrt {-g} = 1$ arises from a serious misunderstanding. The latter condition does not constitute the genuine unimodular constraint; instead, it merely indicates a specific coordinate choice. Covariance of the field equations (arising from tautological diffeomorphism invariance) trivializes the role of the unimodular constraint, as has been shown in the discussion in section \ref{Sec_UGsol}. As we have reiterated multiple times in the manuscript, the accurate UG constraint is precisely the one presented in Eq. \eqref{constraint0}.

Finally, we have discussed the general notion of gauge freedom, as it applies to perturbation theory in General Relativity and related theories, and to cosmology, in particular. We have explained the mistake that underlies claims that UG limits the choice of gauge one might use in cosmology, especially in the treatment of perturbations. In particular, there is a common misconception in the literature arising from the incorrect ``UG constraint'' derived from $\delta \sqrt{-g} = 0$. This misunderstanding leads to the argument that gauge freedom is restricted to choices ensuring the validity of Eq. \eqref{constraintbranden}. For example, according to such an assertion the Newtonian and synchronous gauges would not lead to valid analysis when dealing with cosmological perturbation theory in UG. However, as we have shown in Section \ref{Sec54}, this argument is simply incorrect. In fact, the correct unimodular condition is defined by Eq. \eqref{constraint0}, which places no restrictions on the choice of gauge for perturbation analysis within UG.

We hope that this manuscript will contribute to clarify some of the misconceptions that often arise in discussions about this subject, in general, and its application to UG in cosmology, in particular.



\acknowledgments

G.R.B. is supported by CONICET (Argentina) and he acknowledges support from grant PIP 112-2021-0100225-CO of CONICET (Argentina). G.L. is supported by CONICET (Argentina), and also acknowledges support from the following project grants: Universidad Nacional de La Plata I+D  G175 and PIP 112-2020-0100729-CO  of CONICET (Argentina). AP thanks the support of the ID\# 62312 grant from the John Templeton Foundation, as part of the \href{https://www.templeton.org/grant/the-quantum-information-structure-of-spacetime-qiss-second-phase}{`The Quantum Information Structure of Spacetime' Project (QISS)}. The opinions expressed in this project/publication are those of the author(s) and do not necessarily reflect the views of the John Templeton Foundation. D.S. acknowledges partial financial support from CONAHCYT (M\'exico) project 140630.



\bibliography{bibliografia}
\bibliographystyle{JHEP} 
%
%
%



\end{document}